\documentclass[letterpaper,twocolumn,10pt]{article}
    \usepackage{usenix,epsfig,endnotes}
    \usepackage[inline]{enumitem}
    \usepackage{amsmath}
    \usepackage{url}
    \usepackage{booktabs}
	\usepackage{wasysym}
	\usepackage{makecell}
	\usepackage[table]{xcolor}
    \usepackage[page]{appendix}
    \usepackage{enumitem}
    \usepackage[explicit]{titlesec}

\setlist{parsep=0pt}
\setlist{itemsep=0pt}
\titlespacing{\paragraph}{0pt}{6pt}{0.8em}[]

\begin{document}

\date{}

\title{\Large \bf Partisan: Enabling Cloud-Scale Erlang Applications}

\author{
{\rm Christopher S.\ Meiklejohn} \\
Universit\'e catholique de Louvain \\
Instituto Superior T\'ecnico
\and 
{\rm Heather Miller} \\
Northeastern University \\
\'Ecole Polytechnique F\'ed\'erale de Lausanne
} 

\maketitle

\thispagestyle{empty}

\subsection*{Abstract}

In this work, we present an alternative distribution layer for Erlang,
named \textit{Partisan.} Partisan is a topology-agnostic distributed
programming model and distribution layer that supports several network topologies for different
application scenarios: full mesh, peer-to-peer, client-server, and
publish-subscribe. Partisan allows application developers to specify the
network topology at runtime, rather than encoding topology-specific concerns
into application code. Partisan additionally adds support for more channels, enabling 
users to distribute messages over multiple channels, sometimes in parallel. 

We implement and evaluate Partisan in the Erlang programming language and use it in the evaluation of three scenarios. 
The first scenario compares the raw performance between Distributed Erlang and Partisan, and shows that Partisan performs on par with or better than Distributed Erlang.
The second scenario demonstrates that distributing traffic over multiple connections enables Partisan to perform up to 18x better under normal conditions, and up to 30x better in situations with network congestion and high concurrency.
The third scenario demonstrates, using existing applications, that configuring the
topology at runtime allows applications to perform up to 13.5x better or scale to clusters of thousands of nodes over the general-purpose runtime distribution layer.

\section{Introduction}\label{sec:introduction}
Building cloud-scale distributed applications is becoming increasingly
commonplace. Once a restricted domain for either
scientific computing or data warehousing applications, distributed
applications are now pervasive. Examples of modern distributed applications are: 

\vspace{-3mm}
\begin{itemize}[noitemsep]
	\item distributed databases or infrastructure components that communicate using the \textit{full mesh model};
	\item rich-web or mobile applications that communicate using the \textit{client-server model};
	\item peer-to-peer applications that communicate with other nodes using the \textit{peer-to-peer model};
	\item Internet of Things applications that send data to and receive data from a data center location using the \textit{publish-subscribe model.}
\end{itemize}

\vspace{-2mm}

Despite the pervasiveness of distributed applications, runtime support for building cloud-scale distributed applications remains rare, requiring application developers to build and maintain a communications framework in addition to their application code.

While not yet the norm in industry, there are some notable counterexamples, all of which are
implementations of a distributed actor model; for example: Akka
Cluster~\cite{akkacluster}, Microsoft Orleans~\cite{bykov2011orleans}, and
Distributed Erlang~\cite{wikstrom1994distributed}. Each of these frameworks
enables transparent distributed programming for the platforms they are
designed for, but all three optimize for a single type of application:
low-latency, small-object messaging between nodes in a single cluster,
operating inside the data center, using the \textit{full mesh model.}

However, the modern examples of distributed applications enumerated above show that a single topology is insufficient for the various types of
cloud-scale applications that are being written today. 



In this paper, we present the design of {\em Partisan}, a distributed programming model and distribution layer for Erlang that is meant to be used as an alternative to Distributed Erlang. 
Partisan introduces two important improvements over Distributed Erlang; (1) the addition of multiple runtime-selectable cluster topologies, and (2) the ability to gain additional parallelism by distributing messages over multiple communication channels.

Applications that are developed using the Partisan programming model can specify the cluster topology at runtime. This runtime selection allows applications to choose the most efficient topology for the application at hand without having to modify application code.

Partisan's default topology resembles the default ({\em full mesh}) topology of Distributed Erlang. However, unlike Distributed Erlang, Partisan can distribute traffic over multiple connections to avoid congestion problems observed in Distributed Erlang.

%
%
%
%
%

As Distributed Erlang is general purpose, it can't perform efficiently for all application scenarios. We consider two application scenarios; (1) a distributed database that deals with large objects on smaller clusters (10s of nodes), and (2) a lightweight replicated key-value store for mobile applications that runs on large clusters (100-1000s of nodes).

By leveraging communication channels, we demonstrate up to a 30x improvement on point-to-point messaging, as well as an 13.5x improvement on the distributed database application. By enabling application developers to specify the topology at runtime, we demonstrate the ability to scale the lightweight key-value store application from a cluster of 256 nodes to a cluster of 1024 nodes. 

The contributions of this paper are the following:

\begin{itemize}

\item the design of the Partisan programming model that supports the runtime specification of multiple cluster topologies;

\item the design of the channel-based full mesh backend that enables greater parallelism than possible in Distributed Erlang;

\item an open-source implementation of Partisan that supports five cluster topologies; and

\item a detailed evaluation of Partisan demonstrating increased parallelism through the use of multiple communication channels and increased scalability by specializing the topology to the application at runtime.

\end{itemize}

\section{Preliminaries: Distributed Erlang}\label{sec:background}
Erlang~\cite{armstrong2003making} is a general purpose, concurrent,
functional programming language developed by Ericsson in 1986 for the
construction of highly-available, fault-tolerant concurrent telephony
applications. Erlang has seen much success in industry: Ericsson's AXD301 ATM
switch, WhatsApp's mobile chat application, and the distributed database Riak
(used by the UK's NHS.)~\cite{ghaffari2014investigating}

Erlang applications are constructed using lightweight processes that
communicate with one another. These processes do not share memory: they are
strongly isolated and communicate with one another only through asynchronous
message passing. When one node sends another a message using a primitive
operation \texttt{send} (or $!$), it is delivered into the receiving
processes mailbox (ie.\ queue) and a primitive operation \texttt{receive} is
used to remove a message from the mailbox and handle it in application code.
Processes in Erlang are identified by process identifiers, which can be sent
as messages themselves. Erlang additionally provides functionality that
allows processes to monitor other processes and be notified either when a
process crashes or exits normally by delivering a message to the monitoring
process. Programs in Erlang are written in functional-style, using
single-assignment variables with pattern matching.

Distributed Erlang is an extension that supports transparent Erlang
programming within a cluster of nodes. Several industry products rely on
Distributed Erlang, with the largest known Distributed Erlang cluster being
operated by Ericsson at 200 nodes.\footnote{Personal communication with
author.} Distributed Erlang, by default, establishes a full mesh for
connectivity: when a new node joins the cluster by establishing a connection
to a node already in the cluster, it will also establish connections to all
of the nodes known by its peer, ensuring full connectivity between all of the
nodes. 

\section{Partisan}\label{sec:partisan}
Partisan is a distributed programming model and distribution layer that is
realized as an Erlang library. aimed at providing cloud-scale Erlang
applications. Partisan is meant to be used in lieu of Distributed Erlang to
enable the development of cloud-scale distributed Erlang applications.
Partisan exposes functions that support asynchronous programming regardless
of the topology being used, therefore allowing the developer to alter the
topology during development or at deployment time.

\subsection{Programming Model}
Partisan's programming model provides two sets of operations: membership
operations, that are used for joining and removing Erlang nodes from the
cluster; and messaging operations, that are used for asynchronously
delivering messages between Erlang nodes in the cluster.

Partisan's programming model is designed to be topology-agnostic and
asynchronous. Therefore, all operations in Partisan return immediately and
have backend-specific behavior.  For example, when joining a node in full mesh mode,
the node must be connected to every other node in the cluster; when joining a
node in the client-server mode, if the node is a client, the node will be
redirected to a server node for the connection.

Messaging in Partisan is asynchronous and best-effort: in full mesh
mode, messages will be directly sent to the Erlang node; in peer-to-peer
mode, a message may have to be forwarded through several nodes to reach its
destination based on what connections exist in the cluster.

\begin{table*}[ht!]
\centering
\begin{tabular}{l l l}
\toprule 
 Functionality & Partisan API & Equivalent Distributed Erlang API\\ 
 \midrule 
 Join node to cluster & \texttt{join(Node)} & \texttt{net\_kernel:connect(Node)} \\
 \arrayrulecolor{black!10}\midrule
     \makecell[l]{Remove node from\\ the cluster} & \texttt{leave(Node)} & \texttt{net\_kernel:stop()} \\
 \arrayrulecolor{black!10}\midrule     
     \makecell[l]{Return locally known\\ members of the cluster} & \texttt{members()} & \texttt{nodes()} \\
 \arrayrulecolor{black!10}\midrule     
     \makecell[l]{Forward message\\ asynchronously} & \makecell[l]{\texttt{forward\char`_message(Node,}     
     \\\texttt{~~~~~~~~~~~~~~~~Channel,}
     \\\texttt{~~~~~~~~~~~~~~~~RemotePid,}
     \\\texttt{~~~~~~~~~~~~~~~~Message,} 
     \\\texttt{~~~~~~~~~~~~~~~~Options)}} & \makecell[l]{\texttt{erlang:send(RemotePid,} \\\texttt{~~~~~~~~~~~~Message)}} \\
 \arrayrulecolor{black!10}\midrule     
     \makecell[l]{Forward message\\ asynchronously\\ to gen\_server} & \makecell[l]{\texttt{cast\char`_message(Node,}     
     \\\texttt{~~~~~~~~~~~~~Channel,}
     \\\texttt{~~~~~~~~~~~~~RemotePid,}
     \\\texttt{~~~~~~~~~~~~~Message,} 
     \\\texttt{~~~~~~~~~~~~~Options)}} & \makecell[l]{\texttt{gen\char`_server:cast(ServerRef,}
     \\\texttt{~~~~~~~~~~~~~~~~Message)}} \\

 \arrayrulecolor{black}\bottomrule 
 \end{tabular}
\label{table:api}
\caption{Partisan API}
\end{table*}

\subsection{API}
We now describe the shared API provided by Partisan that is shared across all
of Partisan's membership backend modules.

\begin{itemize}

\item \textbf{Join.} Join a node to the cluster.
This call simulates the \texttt{net\_kernel:connect} functionality.

\item \textbf{Self Leave.} Explicitly leave the cluster.
This call simulates the \texttt{net\_kernel:stop}
functionality.

\item \textbf{Leave.} Explicitly have a node leave the cluster. This
call, when executed at a node in the cluster, will cause \texttt{Node}
to invoke self leave.

\item \textbf{Members.} Return cluster members known locally at this node. In
the event that the peer-to-peer membership library is used, this will only be
members that are directly connected; with the full membership backend, this
will be all members of the cluster. This call simulates the 
\texttt{nodes} functionality.

\item \textbf{Forward Message.} Forward a message to a remote node using
best effort delivery. Returns to the caller immediately and attempts to
deliver the message asynchronously. This call simulates the 
\texttt{erlang:send} (\texttt{!}) functionality.

\item \textbf{Cast Message.} Delivery an asynchronous message to a remote
\texttt{gen\_server} using best effort delivery. Returns to the caller
immediately. This call simulates the \texttt{gen\_server:cast} 
functionality.

\end{itemize}

\subsection{Topologies}
Partisan provides several backend modules for different network topologies.
The topology used by Partisan is specified in the application
environment at runtime.

\begin{itemize}

    \item \textit{Static.} In static mode, Partisan will only
    connect to other nodes that have been explicitly configured at the time
    of node deployment time.
    
    \item \textit{Full Mesh.} In full mesh mode, Partisan will ensure all
    nodes in the cluster are fully connected; in that, each node will connect
    to every other node in the cluster directly, ensuring each node has full
    knowledge of the entire cluster. This topology is an implementation of
    the default configuration of Distributed Erlang.

    \item \textit{Client-Server.} In client-server mode, Partisan will ensure
    that all nodes tagged as clients only connect to nodes tagged as server;
    and all nodes tagged as server nodes will connect to one another.
    Client-server is an implementation of the traditional topology used by
    rich-web and mobile applications.

    \item \textit{Peer-to-Peer.} In peer-to-peer mode, Partisan will have all
    clients connect to one other client in the system and the resulting
    network will approximate an Erd\"os-R\'enyi~\cite{erdHos1964strength,
    erdos1960evolution} model.

    \item \textit{Publish-Subscribe.} In publish-subscribe mode, Partisan
    will connect to preconfigured AMQP~\cite{vinoski2006advanced} message
    broker for node-to-node messaging and dissemination of cluster membership
    information.

\end{itemize}

\subsubsection{Static Membership}
Partisan's static membership backend assumes that nodes participating in the
system will specify the nodes that they wish to connect to at deployment
time: these nodes are specified in a configuration file, or in source code,
and assumes a static network where nodes will not join or leave. This backend
module is primarily used for testing because it reduces nondeterminism in
the network.

The static membership backend uses a single TCP connection for communication
between each node in the cluster, and the failure detector reports failures
when this connection drops. Static membership operates similarly to the
default Distributed Erlang configuration, with the only restriction that
nodes cannot be added or removed from the cluster during cluster operation.

\subsubsection{Full Mesh Membership}
Partisan's full mesh backend provides similar functionality to what is
provided by the default configuration of Distributed Erlang: connections are
established between all nodes in the cluster using a single TCP connection.
Membership is dynamic: nodes can explicitly join or leave the cluster
whenever they desire.

Partisan extends this traditional behavior with a number of new features to
alleviate head-of-line blocking problems and other performance issues in the
design of both Distributed Erlang and Scalable Distributed Erlang.

\paragraph{Channels.} Partisan's full mesh backend supports multiple
connections between nodes in the cluster using channels. This
allows traffic within a cluster to be classified accordingly, and load
balanced across the multiple connections established between each node.

For each channel, and each peer in the cluster, Partisan maintains a single
TCP connection. When a node wishes to send a message to another node in the
cluster, a channel is optionally specified. If a connection exists for that
channel and that peer, that connection process will be sent a message to be
delivered over the TCP connection. If a connection for the channel does not
exist, or a channel is not specified, a default channel and its associated
TCP connection is used to deliver the message.


\paragraph{Monotonic Channels.} Partisan's full mesh backend allows
these named channels to be classified as monotonic or not. Monotonic channels
have a property where each message sent on the channel will subsume a
previous message on the channel. Monotonic channels are useful for performing
load shedding when a particular channel is overloaded with redundant
messages.


\paragraph{Channel Parallelism.} Partisan has the ability to open multiple
TCP connections per channel. This enables additional parallelism by
dispatching and load balancing traffic across multiple TCP connections for
the same named channel and type of traffic.


\paragraph{Membership.} Membership is tracked at each node using a set
and gossiped~\cite{demers1987epidemic} to other nodes in the cluster.
Connections are automatically established as new nodes join the cluster.
Periodically, each node in the cluster will send a copy of this set to its
peers. Upon receipt of the message, the set will be merged with the node's
local copy of the set. This process will continue until a fixed point is
reached.

\subsubsection{Client-Server Membership}
Partisan's client-server backend assumes that each node in the system is
tagged as either a client or a server. Membership is dynamic, but clients are
only allowed to connect to other nodes tagged as servers; server nodes are
only allowed to connected to other nodes tagged as servers. The client-server
topology resembles the traditional hub-and-spoke topology, and can be
implemented by reusing the full mesh topology, and restricting node
connetions between nodes based on their tags.

\subsubsection{Peer-to-Peer Membership}
Partisan's peer-to-peer backend builds upon the
HyParView~\cite{leitao2007hyparview} membership protocol and the
Plumtree~\cite{leitao2007epidemic} epidemic broadcast protocol, both of which
are Hybrid Gossip protocols, where a two-phase approach is used to pair an
efficient dissemination protocol with a resilient repair protocol used to
ensure the efficient protocol can recover from network partitions.

\paragraph{HyParView.} HyParView is a hybrid gossip algorithm that provides a
resilient membership protocol by using partial views to provide global system
connectivity in a scalable way. Using partial views ensures scalability;
however since each node only sees part of the system, it is possible that
failures of other nodes break connectivity or greatly increase routing
length. To overcome these problems, HyParView uses two different partial
views that are maintained with different strategies. The challenge is to
ensure that the combination of all partial views at all nodes form a single
connected component.



\paragraph{Plumtree.}
Plumtree is a hybrid gossip algorithm that provides reliable broadcast by
combining a deterministic tree-based broadcast protocol with a gossip
protocol. The tree-based protocol constructs and uses a spanning tree to
achieve efficient broadcast. However, it is not resilient to node failures.
The gossip protocol is able to repair the tree when node failures occur. Thus
the Plumtree protocol combines the efficiency of spanning trees with the
resilience of gossip.



\paragraph{Transitive Delivery.} In a HyParView cluster, nodes may want to
message other nodes that are not directly connected. To maintain the existing
semantics of Distributed Erlang, Partisan needs a mechanism to support
messaging between any two nodes in a cluster.

To achieve this, Partisan's peer-to-peer membership backend uses an instance
of the Plumtree protocol to compute a spanning tree rooted at each node.
Periodically, using a configurable interval, Partisan will broadcast a
heartbeat message with a timestamp using Plumtree to ensure the tree is
maintained; in the event the tree is disconnected, the normal Plumtree repair
process is used. When attempting to send to a node that is not directly
connected, the spanning tree is used to forward the message down the leaves
of the tree in a best-effort method for delivering the message to the desired
node. This is similar to the approach taken by
Cimbiosys~\cite{ramasubramanian2009cimbiosys} to prevent livelocks in their
anti-entropy system.

\subsubsection{Publish-Subscribe Membership}
Partisan's publish-subscribe backend builds upon the Advanced Message
Queueing Protocol (AMQP) standard. AMQP is a wire-level protocol, and
therefore only specifies the format messages should take. This allow Partisan
to operate on top of arbitrary backends that support the AMQP standard, such
as cloud-based offerings like Amazon's Simple Queue Service, Google's Cloud
Pub/Sub and Microsoft's Azure Service Bus, and local, on-premise solutions
like RabbitMQ.

Partisan's publish-subscribe backend also only establishes outbound
connections from nodes for bidirectional messaging, which makes it ideal for
use in environments where outbound communication is prohibited, such as
Amazon's Lambda and Google's Cloud Functions.

When using the publish-subscribe backend, a single queue is used for 
dissemination of membership information, that is subscribed to by all nodes
participating in the system.  For each Erlang node in the membership, a queue 
is registered for messages destined for that node; each Erlang node subscribes 
to its own channel.

\subsection{Design Considerations}
In order to provide a topology agnostic programming model, we do not support
features of Distributed Erlang that are unable to be supported across all
topologies.

\paragraph{Remote Monitoring}
Monitors in Erlang allow processes to be notified when other processes
terminate. Remote monitoring is straightforward: as long as a connection
remains open to the remote host where the process being monitored is
executing, the monitor will operate correctly. However, if the connection to
the remote host is lost, regardless of process state, the monitor will report
the process as terminated.

Remote monitoring introduces a number of complications, as it is only
possible if the node where the remote process is executing is directly
connected. This is further complicated because of the topologies Partisan
supports:

\begin{itemize}

    \item \textit{Client-server backend.} Client nodes only know about one or
    more servers, and therefore cannot remotely monitor processes on other
    client nodes.

    \item \textit{Peer-to-peer backend.} Nodes are only partially connected,
    the remote processes may not be exceuting on one of the connected nodes.
    The only alternative would be to directly connect that node, causing the
    cluster to reorganize. 
    

    \item \textit{Publish-subscribe backend.} Remote monitoring is not
    possible given no connections are maintained directly between nodes.

\end{itemize}

If remote monitoring is required, Partisan can additionally connect nodes over 
Distributed Erlang to provide this functionality.

\paragraph{Synchronous Invocations} 
The generic server abstraction \texttt{gen\_server} contains two methods for
making calls: \texttt{cast}, for asynchronous invocation, and \texttt{call},
for synchronous invocation. \texttt{call}, for synchronous invocation relies
on the use of a monitor, whereas \texttt{cast}, for asynchronous invocation,
does not. When a call is made, a monitor is is placed on the
\texttt{gen\_server} process that the call is being made to, and if that
process dies, the call returns with an error code instead of blocking and
waiting for a response indefinitely. A similar issue exists for the generic
finite state machine, \texttt{gen\_fsm}. Since these calls rely on the use of
a remote monitor, these calls are unsupported by default.


\paragraph{Process Identifiers}
Process identifiers in Distributed Erlang combine a unique node identifier
with process identifier to identify the process globally. The node identifier
is encoded as an integer, and is relative to the node the process identifier
is being viewed from. As local processes always have the node identifier of
0, when process identifiers are transmitted between nodes, the process
identifiers are translated based on the receiving nodes membership view. 


Supporting process identifiers in Partisan, without changing the internal
implementation of Erlang's process identifiers, is not possible without
allowing nodes to directly connect to every other node. Instead of relying on
Erlang's process identifiers, Partisan recommends that processes that wish to
receive messages from remote processes locally register a name that can be
used instead of a process identifier when sending the message. We envision
that a future version of Partisan could handle this automatically as part of
message serialization.

\paragraph{Message Ordering}
Distributed Erlang provides unreliable FIFO delivery between any two sending
processes~\cite{svensson2007programming}. This means that, given two
processes, the receiver will always receive messages from the sender in
sending order; however, groups of messages may be omitted, as long as
ordering is preserved.

Partisan provides best-effort ordering, depending on the topology and
configuration of that topology.  Given two peers, we make the following 
guarantees:

\begin{itemize}

    \item \textit{Full mesh backend.} With a single connection between peers,
    ordering is preserved between reconnections. With multiple channels,
    ordering is preserved per channel between reconnections. With multiple
    connections per channel, ordering is only preserved, between
    reconnections, if a routing partition key for the sender is provided; no
    guarantees are provided under random partitioning.

    \item \textit{Client-server backend.} Same as the \textit{full mesh}
    model, only ordering is only preserved between client and server nodes.

    \item \textit{Peer-to-peer backend.} As messages may take any path to
    reach a recipient, FIFO is not guaranteed.

    \item \textit{Publish-subscribe backend.} FIFO is guaranteed for the
    lifetime of the broker and exchange.

\end{itemize}

\vspace{-4mm}

\subsection{Implementation}
Partisan is implemented as a library for Erlang 19.3 and requires no
modifications to either the compiler or VM. It is implemented in 6.7 KLOC and
is available as open source on GitHub~\cite{partisan-repository}. The open
source implementation of Partisan has several industry adopters.

Cluster topologies are Erlang modules
that implement the \texttt{partisan\_peer\_service\_manager} behavior. Users
can implement their own topologies by providing a module that implements this
behavior. Client applications interact with the Partisan system through the
\texttt{partisan\_peer\_service} module, which exposes the API presented in
Table~\ref{table:api}.

Monotonic channels are implemented as Erlang
processes that receive messages to be sent on the network to another node in
the system. Whenever a monotonic channel process receives a message to be
delivered over the channel, if the process's mailbox contains more than 1
message in the queue, the message is dropped. These messages are only dropped
within a particular window: the system will ensure that at least 1 message is
sent within a particular window, to ensure progress.

Our implementation uses the following optimizations:

\textit{Binary Serialization.} Serialization to Erlang's external term format
occurs inside the Erlang VM, and in Distributed Erlang, can maximize sharing
of the underlying data structures before transmitting the data structure on
the wire. However, since serialization is invoked outside the VM in Partisan,
a one-time binary object is generated off-heap and immediately dereferenced
once the object is transmitted. Therefore, we cannot take advantage of
reusing existing, shared structures. Using a technique from
Thompson~\cite{vagabond}, recursive terms are encoded as lists and base types
encoded as binaries before transmission to maximize binary reuse. This
serializer to users implementing their own backend.

\textit{Overflow of LISTEN Queue.} When building large clusters
using the \textit{full mesh} backend, the TCP LISTEN queue can overflow when other
members of the cluster establish new connections. For example, a cluster
of $N$ nodes, when growing to a cluster of $N + 1$ nodes will cause the
cluster to establish $N$ new connections to the joining node. This is
exacerbated when using both the channel and parallelism features
of the full mesh backend; causing a joining node to receive $N * C * P$
inbound connections at the same moment, overflowing the
node's LISTEN queue, causing timeouts. To
mitigate this, when existing nodes learn about a joining node they should
connect to, only establish a single connection to a joining node every
refresh interval.

\textit{Connection Cache.} To avoid any unnecessary contention when sending
messages, a cache (implemented as an ETS table) is used to store the list of
open connections. ETS (Erlang Term Storage) tables are processes that manage
shared memory storage tables in the VM that can be concurrently accessed by
multiple processes for reading. This cache is available to users implementing
their own backend.

\begin{figure*}[ht!]
    \includegraphics[width=\textwidth]{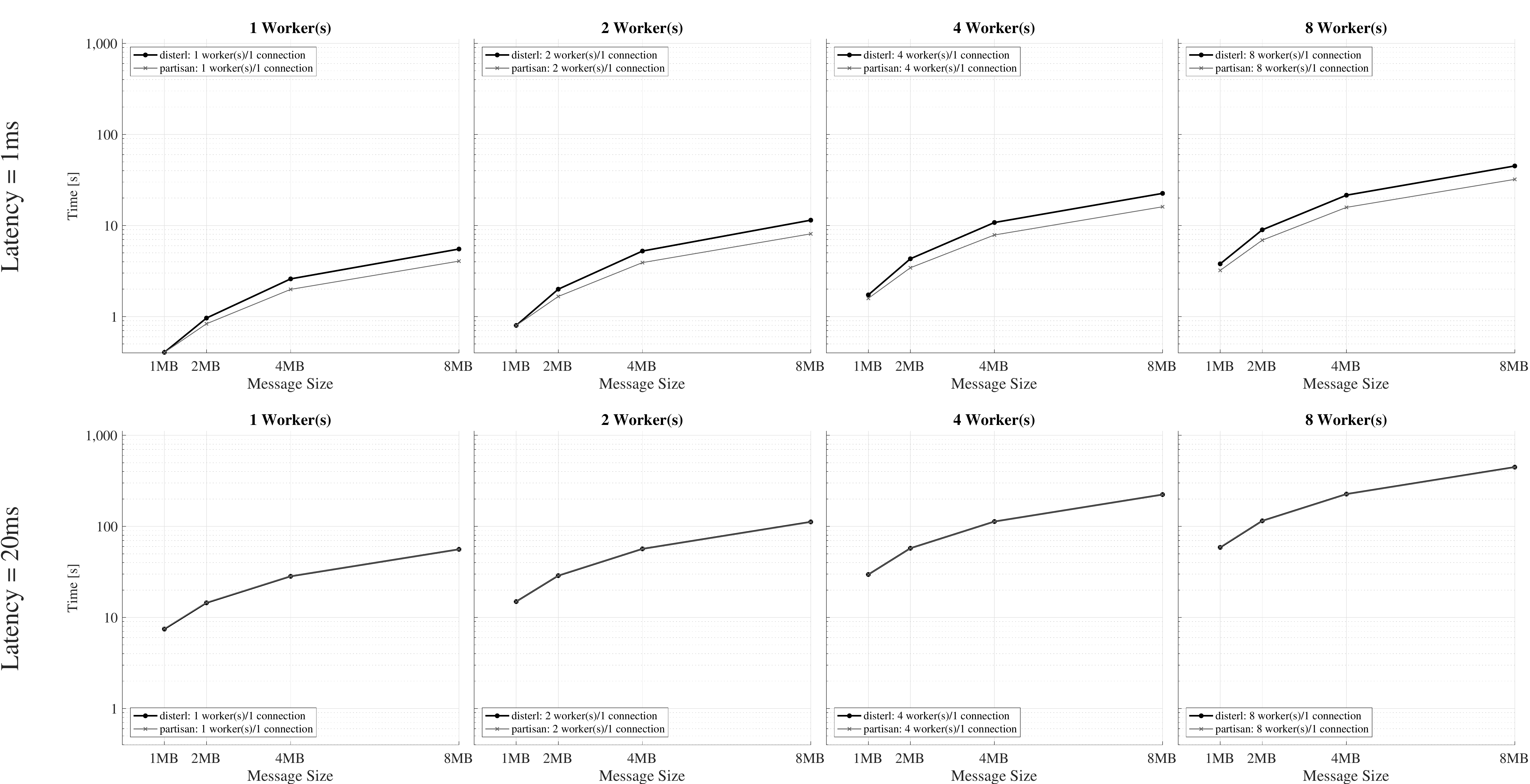}
    \caption{Performance of 1 channel Partisan vs. Distributed Erlang.}
    \label{fig:unicast}
\end{figure*}

\section{Evaluation}

To evaluate the design decisions behind Partisan, and to validate its implementation, we focus on providing answers to the following overarching questions:

\begin{itemize}

	\item Can a distribution layer removed from the Erlang VM perform on-par or better than one provided within the Erlang VM?
	
	\item Is it advantageous to separate different kinds of messages into dedicated channels? If so, what level of parallelism is best? 
	
	\item Can the runtime selection of a network topology offer better performance than a general-purpose topology that provided by the system implementation?
	
	\item How do realistic applications, including one on a large cluster, perform atop of the Partisan distribution layer?

\end{itemize}

%
%
%
%

We answer these questions in the following subsections.

In Section~\ref{sec:dist-as-a-library}, we demonstrate that a library-based implementation of a distribution layer can outperform a general-purpose distribution layer provided the runtime. We evaluate the performance of both Distributed Erlang and Partisan under different network latencies and varying workloads.

In Section~\ref{sec:dist-messages-channels}, we examine the benefits of channels. We demonstrate that using multiple channels can prevent interference between different types of messages. We further demonstrate that parallelizing transmission on such channels can be beneficial when there is either high concurrency or when there is increasing network latency. Both of these design decisions allow us to mitigate the effects of head-of-line blocking. 

To demonstrate that no single topology is sufficient for optimal performance
of all applications, in Section~\ref{sec:one-size-fits-all} we examine two existing Erlang applications. The first,
Riak Core~\cite{klophaus2010riak}, is the underlying infrastructure for the distributed
database Riak~\cite{riak}, the research database Antidote~\cite{antidote},
and several industry products~\cite{goldman, projectfifo, nkbase}. The second, is
the research language, Lasp~\cite{meiklejohn2015lasp}, designed for
large-scale, coordination-free programming. 

\subsection{Distribution as a Library}
\label{sec:dist-as-a-library}

Distributed Erlang is implemented as an extension within the Erlang virtual
machine. Message serialization, connection maintenance, and data transmission
are all handled by mechanisms inside of the virtual machine that minimize
redundant data serialization, and avoid penalties from messages copies
between different processes heaps. As Partisan is implemented as a library,
we set out to evaluate the viability of running a distribution layer that
does not cohabit the virtual machine.

\begin{figure*}[ht!]
    \includegraphics[width=\textwidth]{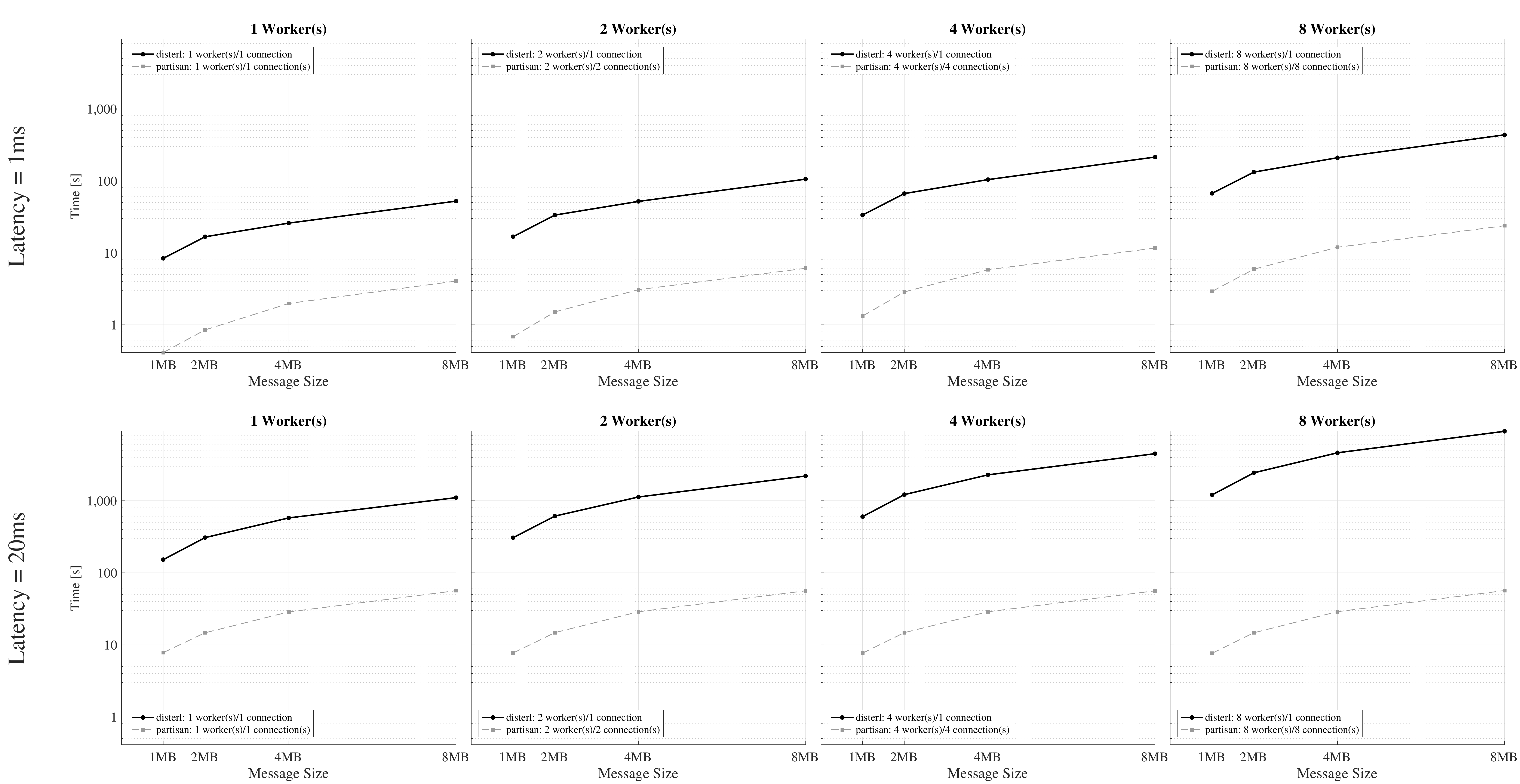}
    \caption{Performance of 3 channel (vnode, metadata, gossip) Partisan vs. Distributed Erlang on a Riak Core cluster.}
    \label{fig:unicast:riakcore}
\end{figure*}

To evaluate this, we ran a two node Erlang cluster, simulating 1ms and 20ms
RTT latencies\footnote{1ms and 20ms are representative of intra-, and inter-
availability zone latentices using Amazon AWS, respectively.}, and recorded
the execution time taken for $N$ processes running on the the second node to
receive 1,000 messages sent $N$ worker processes running on the first node.
Figure~\ref{fig:unicast} demonstrates that Partisan's distribution layer,
implemented in Erlang, can achieve the same performance as the VM-supported
Distributed Erlang, and in some cases, outperform Distributed Erlang.

\vspace{-2mm}

\subsection{Distributing Messages into Channels}
\label{sec:dist-messages-channels}

Riak Core is a distributed programming framework in Erlang based on the
Amazon Dynamo~\cite{decandia2007dynamo} model. In Dynamo, a distributed hash table is used to route
requests among nodes in a cluster. The hash space is broken into a set of
partitions, and distributed to a fixed set of virtual nodes, each of which is
claimed by a node using a claim algorithm, then stored in a data structure,
known as the ring. Requests are routed using consistent hashing, which
minimizes the impact of reshuffling when nodes join or leave the cluster.


Background processes, such as Riak Core's metadata anti-entropy and Riak
Core's ring gossip, can interfere with messages in the request path, as they
transmit large objects on the same channel as requests, and can create
interference, such as head-of-line blocking problems. To examine the effect
of this, we ran the same unicast benchmark on a 3 node Riak Core cluster
comparing Distributed Erlang and Partisan. Partisan was configured to
distribute traffic across 3 channels: request traffic, metadata
anti-entropy traffic, and ring gossip.

Figure~\ref{fig:unicast:riakcore} demonstrates that performance of Partisan,
when moving background activities to seperate channels, is at best, $12.5$
times faster, averaging an order of magnitude of improvement.

\begin{figure}[h!]
    \includegraphics[scale=.38]{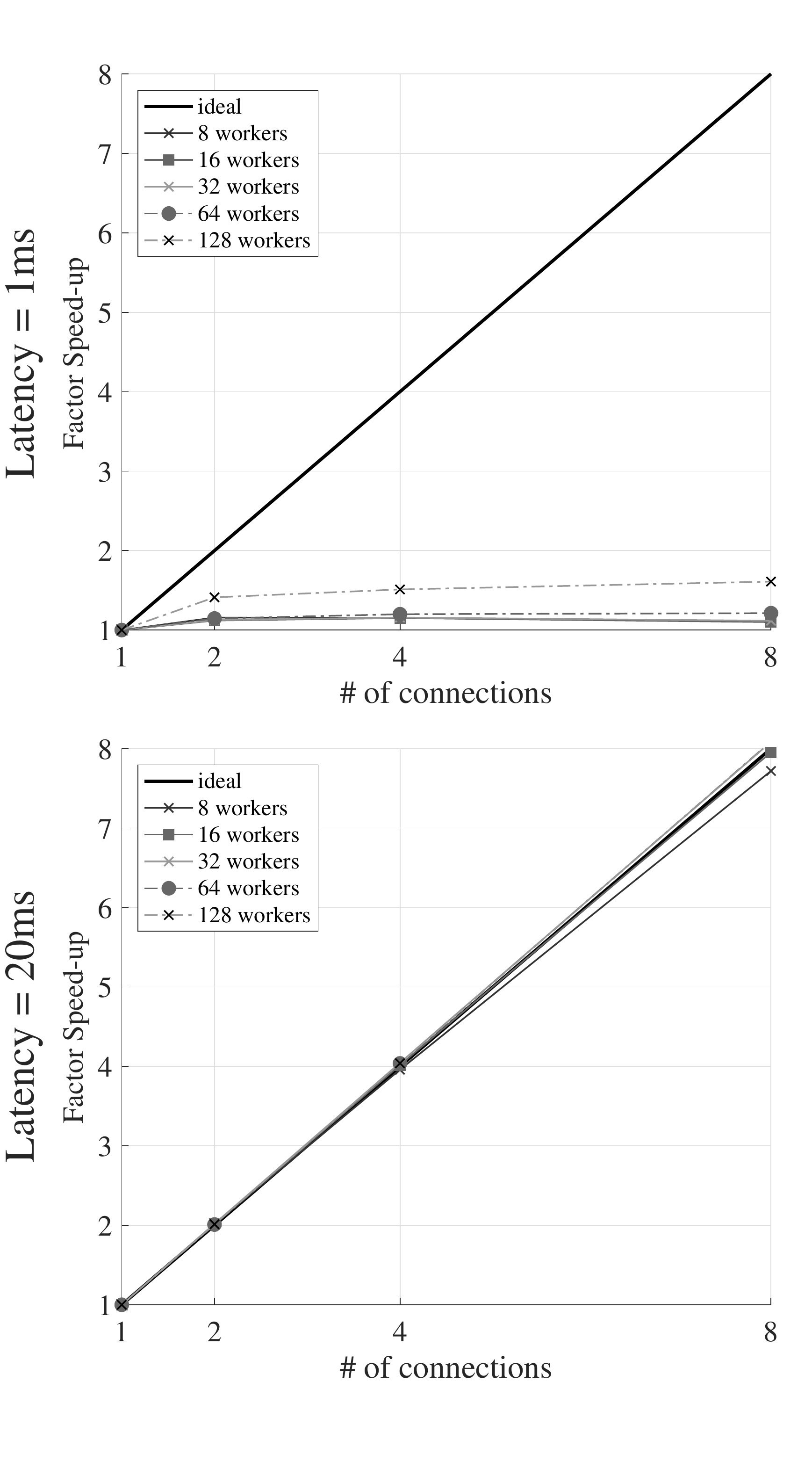}
    \vspace{-6mm}
    \caption{Performance of different parallel Partisan configurations with varying latency with a 1MB payload on a Riak Core cluster.}
    \vspace{-2mm}
    \label{fig:unicast:parallel:scaling}
\end{figure}

\begin{figure}[h!]
    \includegraphics[width=8cm]{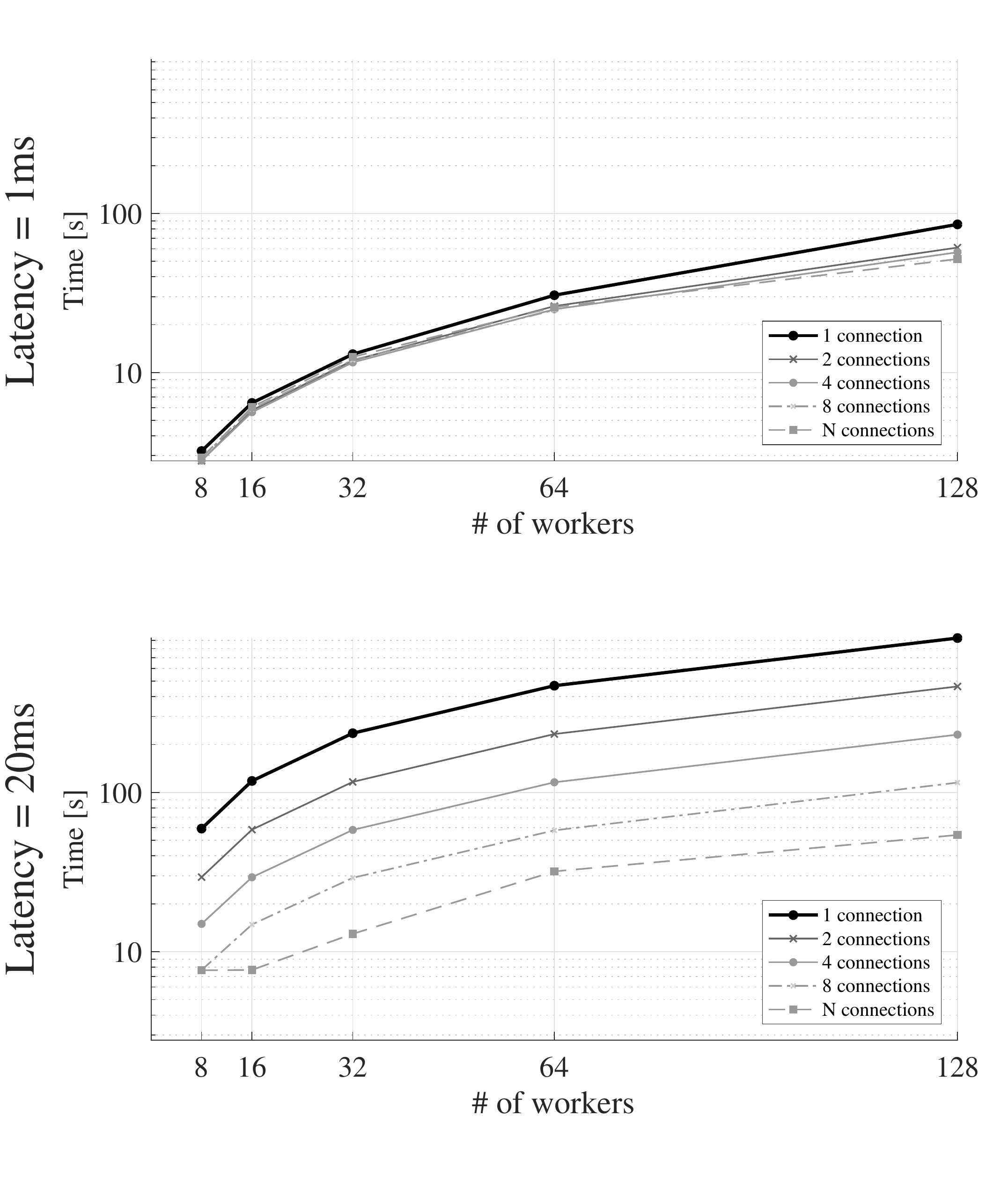}
    \vspace{-6mm}
    \caption{Performance of parallel Partisan configurations vs. Distributed Erlang for 1MB payload.}
    \vspace{-2mm}
    \label{fig:unicast:parallel}
\end{figure}

Under increasing network latency or increasing concurrency, Partisan can
leverage additional connections per channel to scale near linearly.
Figure~\ref{fig:unicast:parallel:scaling} demonstrates the same unicast
benchmark under different levels of concurrency, by increasing the number of
workers while maintaining a fixed set of 8 connections for a single channel
with a 1MB payload. However, under low concurrency, when the cost of sending
the message is minimal, the additional connections are less beneficial and
mostly add overhead to the Erlang scheduler.

Each sender has affinity to a particular connection, using the workers's
process identifier to determine which connection to use to route it's traffic
to the destination. Figure~\ref{fig:unicast:parallel} demonstrates the
benefits from leveraging additional connections in different latency
configurations for a 1MB payload.

\subsection{No ``One Size Fits All'' Topology}
\label{sec:one-size-fits-all}

To motivate the use of multiple topologies, we implement two applications in Riak
Core, and one large-scale application in Lasp.

\subsubsection{Riak Core}

Our first application is a simple echo service, implemented on a 3
node Riak Core cluster. For each request, we generate a binary object, uniformly
select a partition to send the request to, and wait for a reply containing
the original message before issuing the next request. We use the
aforementioned benchmarking strategy: we record the execution time for $N$
worker processes to issue 1,000 requests from 1 node uniformly to partitions
distributed across the cluster. When there is more than one connection
available per channel, Partisan is configured to partition traffic based on
the partition identifier.

\begin{figure*}[h!]
    \includegraphics[width=\textwidth]{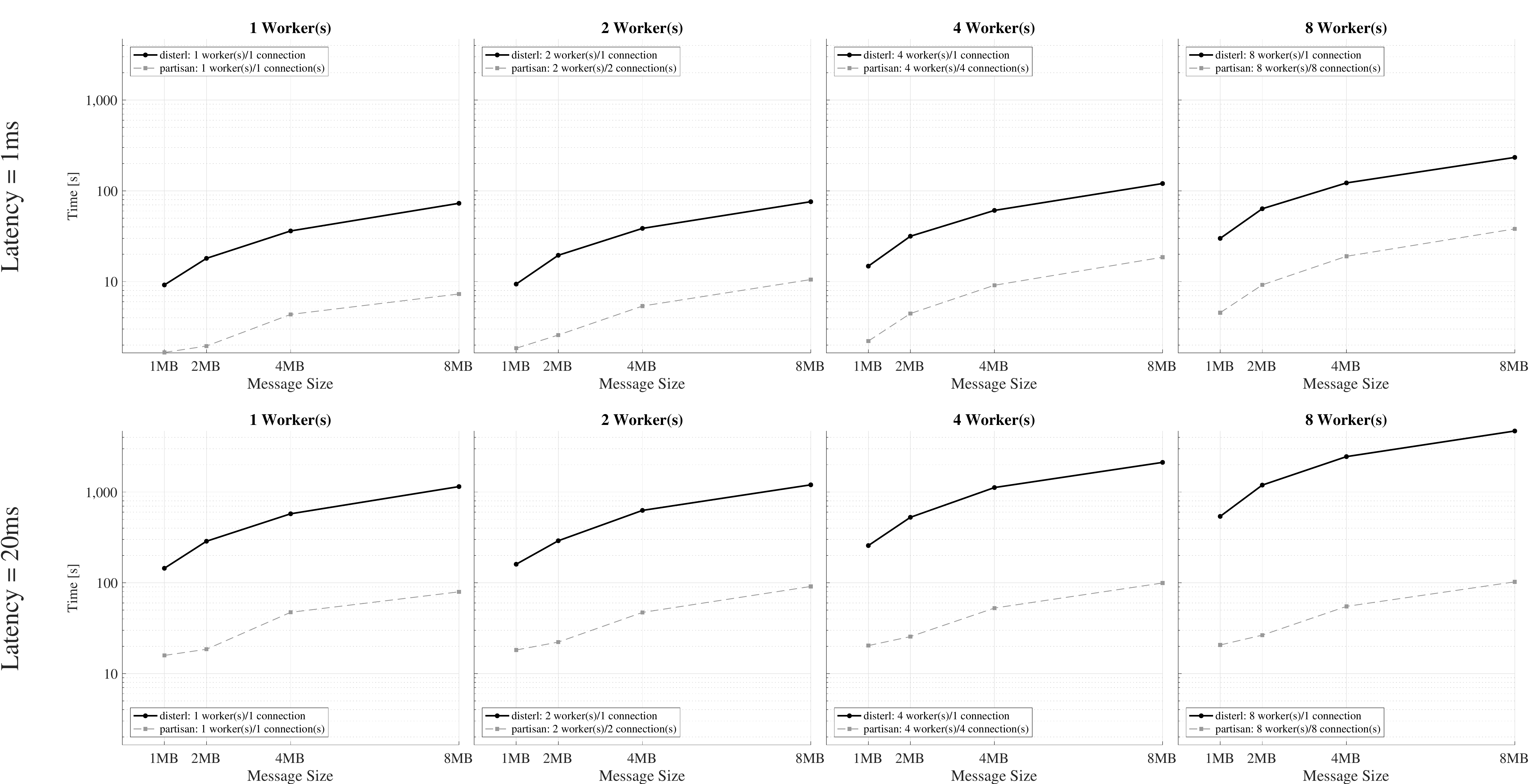}
    \caption{Partian vs. Distributed Erlang: single vnode Echo service implemented in Riak Core.}
    \label{fig:riakcore:echo}
\end{figure*}

Figure~\ref{fig:riakcore:echo} demonstrates that Partisan exhibits an
order of magnitude improvement over Distributed Erlang at low latencies, and
approaches two orders of magnitude as the latency increases. This is due to
the fact that as concurrency and network latency increases, Partisan can more
efficiently leverage the use of additional connections to exploit
parallelism.

\begin{figure*}[h!]
    \includegraphics[width=\textwidth]{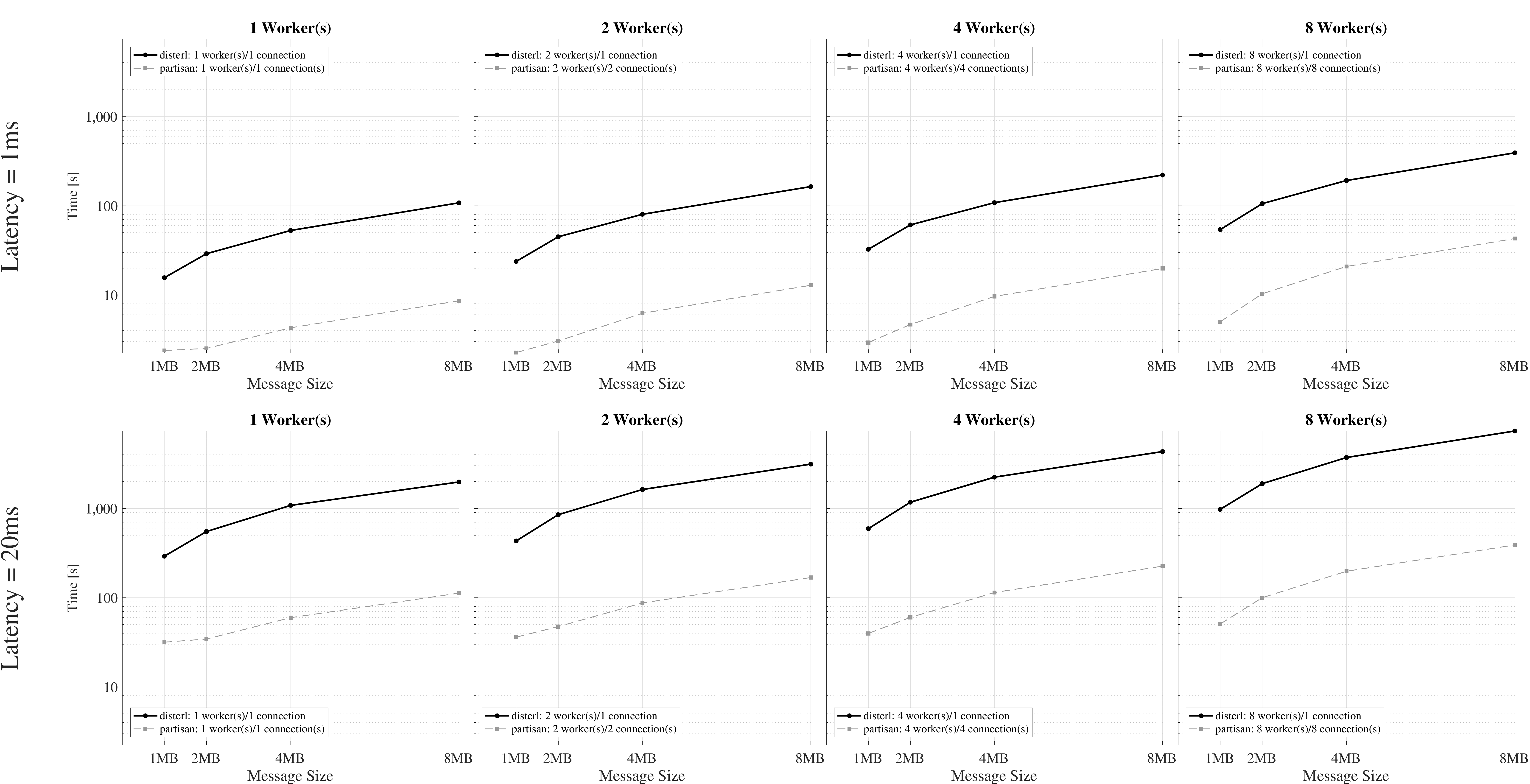}
    \caption{Partisan vs. Distributed Erlang: memory-based KVS implemented in Riak Core.}
    \label{fig:riakcore:fsm}
\end{figure*}

Our second application is a memory-based key-value store, similar to the Riak
database, implemented on a 3 node Riak Core cluster. Each request uses a
quorum intersection request pattern, where get and put requests are issued to
3 partitions, based on where the key is hashed to Riak Core's distributed
hash table (along with its two clockwise neighbors), and the response is
returned to the user once 2 out of the 3 partitions have replied. This
pattern involves multiple nodes in the request path, and each partition
simulates a 1ms storage delay\footnote{1ms is representative of a sequential
seek of 1MB of data.} in the request path. We reuse the aforementioned
benchmarking strategy: we record the execution time for $N$ worker processes
issue 1,000 requests from 1 node uniformly to partitions distributed across
the cluster using Riak Core's claim algorithm. We vary the workload 1:1
between get and put operations, with selecting keys from a normal
distribution of 10,000 keys, with an object payload of 1MB. Similar to
Figure~\ref{fig:riakcore:echo}, Figure~\ref{fig:riakcore:fsm} also
demonstrates that on a KV workload, Partisan can achieve an order of
magnitude improvement over Distributed Erlang.

\vspace{-2mm}

\subsubsection{Lasp}

Lasp is a programming model designed for large-scale coordination-free
programming. Applications in Lasp are written using shared state: this shared
state is stored in an underlying key-value store, and is fully replicated
between all nodes in the system. Applications always modify their own copy of
the shared state, and propagate the effects of their changes to other nodes
in the network. Lasp ensures that applications always converge to the same
result on every node through the use of convergent data structures known as
Conflict-Free Replicated Data Types~\cite{shapiro2011conflict}, combined with
monotone programming~\cite{alvaro2011consistency}.

For our Lasp application, we simulate an advertisement counter, modeled after
the Rovio advertisement counter scenario for Angry Birds, where each client
keeps a replica of distributed counters, incrementing each counter when an
advertisement is displayed. Once a certain number of impressions are reached,
the counter is disabled. The advertisement interval was fixed at 10 seconds,
and the propagation interval for state was fixed at 5 seconds. The total
number of impressions was configured to ensure that the experiment would run
for 30 minutes. We evaluated both \textit{client-server} and
\textit{peer-to-peer} topologies for varying cluster sizes, ranging from 32
all the way up to 1,024 node clusters. For both topologies, we propagate the
full state of the objects in the local store to the nodes's peers at each
propagation interval.

\begin{figure}[t!]
    \includegraphics[scale=.3]{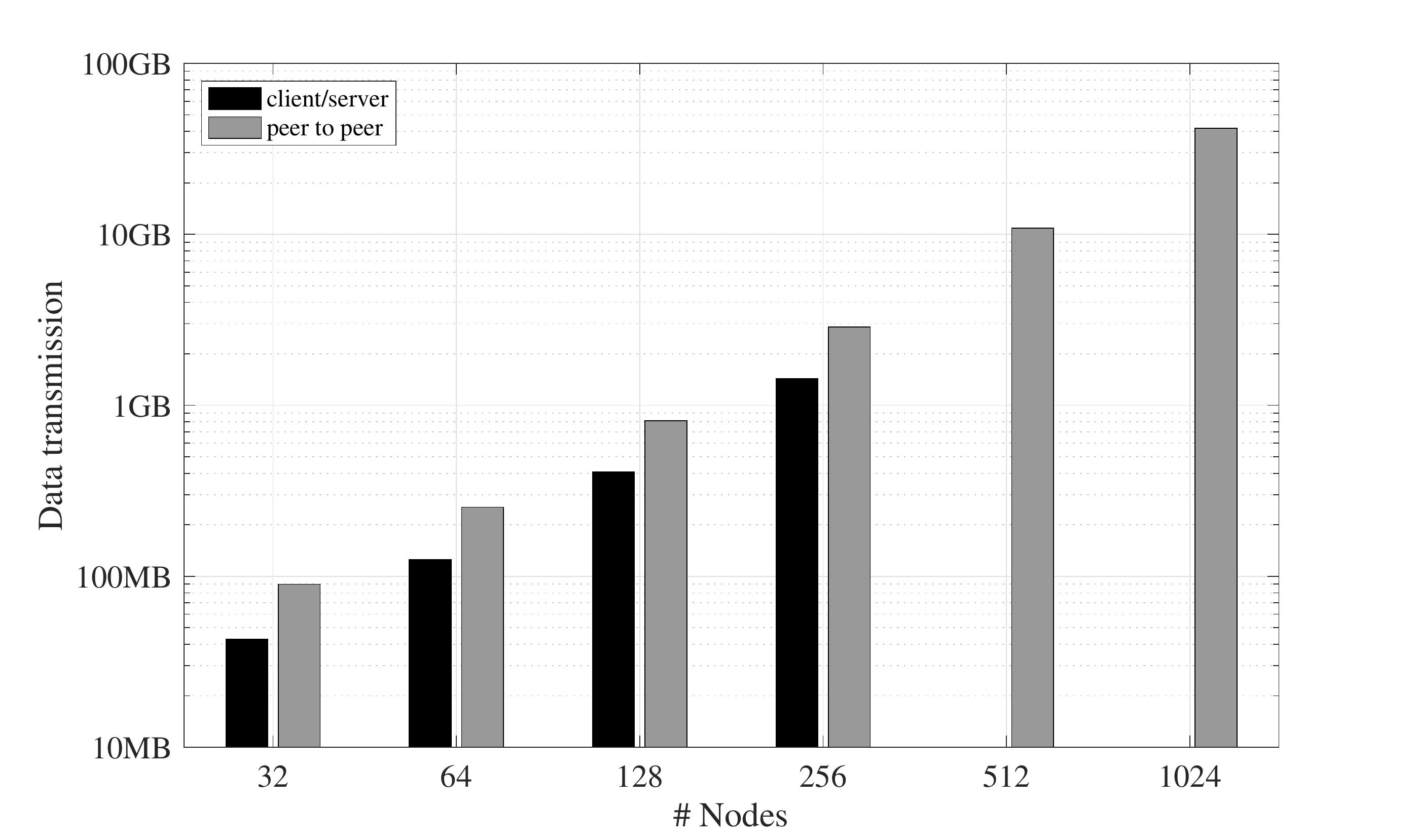}
    \vspace{-6mm}
    \caption{Comparison of data transmission for Lasp deployed on two topologies; client-server and peer-to-peer across various cluster sizes (32 to 1024 nodes).}
    \vspace{-2mm}
    \label{fig:lasp}
\end{figure}

For this evaluation, we used a total of 70 m3.2xlarge
instances in the Amazon EC2 cloud computing environment, within the same
region and availability zone. We used the Apache
Mesos~\cite{hindman2011mesos} cluster computing framework to subdivide each
of these machines into smaller, fully-isolated machines using cgroups. Each
virtual machine, representing a single Lasp node, communicated with other
nodes in the cluster using Partisan.

Figure~\ref{fig:lasp} demonstrates that, in this experiment, the
client-server topology fails to scale above a 256 node cluster whereas the
peer-to-peer topology scales to 1,024 (at which point we encountered issues
with Apache Mesos.) Transmission growth is reported as the total across all
nodes and is impacted by two factors:

\textit{(1) Choice of topology.} The client-server topology has no
redundancy and uses the server as a coordination point; whereas, the
peer-to-peer topology has redundancy introduced as part of the resiliency
of the topology.

\textit{(2) Choice of data structure.} The distributed data structure
used, the G-Counter, grows on the order of the number of clients in the
system.

We refer the readers to~\cite{Meiklejohn:2017:PEL:3131851.3131862} for a
full treatment of the large-scale Lasp evaluation.

\vspace{-2mm}

\section{Related Work}\label{sec:relatedwork}

\vspace{-1mm}

Ghaffari~\cite{ghaffari2014investigating} has identified several factors
that limit the scalability of Distributed Erlang:

\textit{(1) Global Commands.} When $0.01$ percent of commands are
\textit{global} commands, commands that require coordination of all nodes
in the cluster using a mechanism similar to 2PL/2PC, operations can take
up to $20$ seconds, and cluster scalability is limited to $\approx{}60$
nodes.

\textit{(2) Data Size.} Increasing payload sizes of messages between
nodes limits the throughput of the cluster. Ghaffari does not provide
explanation for this, but presumably this problem arises from
head-of-line blocking and the cost of serialization and deserialization.

\textit{(3) Remote Procedure Calls.} Remote Procedure Calls limit
scalability, as each call is serialized through a single server process
that handles all Remote Procedure Calls.  Head-of-line blocking, and 
the maximum throughput of a single process, obviously contribute to 
the scalability problems of scaling RPC in Erlang.

Ghaffari et al.~\cite{ghaffari2013scalable} also identified that Remote
Procedure Call invocations were a limiting factor in Riak 1.1.1's
$\approx{}60$ node limitation on linear scalability, but that as no global
operations were used by the database, was not limited by global operations.

Chechina et al.~\cite{chechina2012design} propose that there are two
fundamental challenges that must be overcome in Distributed Erlang to scale
to hundreds of nodes. Specifically, (i) {\em transitive connection sharing},
and (ii) {\em explicit process placement}. In the case of (i), transitive
connection sharing between all nodes in the cluster requires that each
instance of the Erlang VM maintains data structures quadratic in the number
of nodes in the cluster. In the case of (ii), once clusters grow large
enough, determining where to place computational processes in the network
becomes a challenge to ensure proper supervision, fault-tolerance, and
balanced cluster performance.

The authors propose two solutions, two components of Scalable Distributed
Erlang, to solve these problems:

\textit{(1) Reducing transitive connection sharing.} By subdividing nodes
into smaller groups and only supporting full connectivity within each group
and not across groups, nodes limit the number of nodes that they have to
connect to, perform failure detection on, and replicate the global process
registry of. In this model, each node can become a member of multiple groups
and can explicitly request a connection with another node in the system,
without transitive connection sharing.

\textit{(2) Semi-explicit process placement.} When spawning a new process,
per-node attributes can be used to filter the list of available nodes to
choose from for hosting that process. This allows developers to
target nodes by available memory, or other user-defined attributes.

While these changes enable Scalable Distributed Erlang to break through the
scalability bottleneck with global operations previously identified by
Ghaffari et al.~\cite{ghaffari2013scalable,ghaffari2014investigating},
scaling up to 256 nodes~\cite{chechina2017evaluating}, these solutions still
assume that explicit process naming through the global registry is desirable,
from an application developer point of view. Additionally, a node that
participates in too many groups also will fall into the same trap of
replicating too much information.

Existing distributed actor systems, such as Akka Cluster~\cite{akkacluster}
and Microsoft's Orleans~\cite{bykov2011orleans, bernstein2017geo} share
similar designs to Riak Core. While these systems differ slightly in their
programming models, they both use a distributed hash table for distributing
the placement of actors within the cluster. We believe the techniques in
presented in this paper are applicable to both of these systems, as they have
been demonstrated as applicable to Riak Core.

\vspace{-3mm}

\section{Conclusion}\label{sec:conclusion}

\vspace{-1mm}

We presented Partisan, a distributed programming model and distribution layer
for Erlang that provides the ability for users to specify cluster topologies
at runtime, without requiring modifications to application code. Partisan's
default topology outperforms Distributed Erlang through the use of channels
that can exploit parallelism under high concurrency or increasing latency,
thereby reducing the impact of network interference such as head-of-line
blocking. These design decisions resulted in a 13.5x - 30x performance
improvement for real Erlang applications. As our modifications have
demonstrated performance gains for Riak Core, we believe these design
decisions can lead to improved performance for systems with similar designs,
such as Microsoft Orleans and Akka Cluster.

{\footnotesize \bibliographystyle{acm}
\bibliography{main}}

\end{document}